# Measurement and Analysis of Quality of Service of Mobile Networks in Afghanistan – End User Perspective

M. A. Habibi, M. Ulman, J. Vaněk, J. Pavlík

Faculty of Economics and Management, Czech University of Life Sciences Prague, Czech Republic

## Abstract

Enhanced Quality of Service (QoS) and satisfaction of mobile phone user are major concerns of a service provider. In order to manage network efficiently and to provide enhanced end – to – end Quality of Experience (QoE), operator is expected to measure and analyze QoS from various perspectives and at different relevant points of network. The scope of this paper is measurement and statistically analysis of QoS of mobile networks from end user perspective in Afghanistan. The study is based on primary data collected on random basis from 1,515 mobile phone users of five cellular operators. The paper furthermore proposes adequate technical solutions to mobile operators in order to address existing challenges in the area of QoS and to remain competitive in the market. Based on the result of processed data, considering geographical locations, population and telecom regulations of the government, authors recommend deployment of small cells (SCs), increasing number of regular performance tests, optimal placement of base stations, increasing number of carriers, and high order sectorization as proposed technical solutions.

## Introduction

Existing heterogeneous environment of wireless networks is a complex mixture of Long – Term Evolution (LTE), LTE – Advanced (LTE – A), Wireless Fidelity (Wi-Fi), Universal Mobile Telecommunications System (UMTS), High Speed Packet Access (HSPA) and even General Packet Radio Service (GPRS) and Enhanced Data rates for GSM Evolution (EDGE) (Damnjanovic, et al., 2011). Massive deployment of diverse access technologies improves QoS and increase end user satisfaction but raises many challenges e.g. network complexity, radio resource management, mobility management, etc. (Vondra and Becvar, 2016). On the other hand, end user demand for enhanced quality of data service and multimedia applications is dramatically growing (Olwal, 2016), which is moving researchers, standardization bodies, vendors and operators forward to introduce new techniques and propose various schemes in the area of QoS.

QoS in communication network is the capability of a service provider to provide satisfactory end – to – end service for its end user which includes voice quality of telephony service, signal strength, service availability, low call blocking and dropping probability, minimum delay, high data rates for data service, multimedia applications, etc. The International Telecommunication Union (ITU) has defined QoS as, '*totality of characteristics of a telecommunications service that bear on its ability to satisfy stated and implied needs of the user of the service*' (ITU, 2008). Mobile phone user not only evaluates the performance of a cellular network but considers so called QoE i.e. the price of the service, the perceived quality of both content and the easy use of an application or of the mobile equipment. According to the European Qualinet Community, the QoE is '*the degree of delight or annoyance of the user of an application or service. It results from the fulfillment of his or her expectations with respect to the utility and/or enjoyment of the application or service in the light of the users' personality and current state*' (Callet, et al., 2013).

There is clear difference between QoS and QoE. While QoS deals with performance aspects of physical system, QoE deals with end user assessment of a system. QoS has technology-oriented approach and it relies on analytic approaches and empirical or simulative measurements, but QoE requires a multi-disciplinary and multi-methodological approach for its understanding. It is crucial to remember that QoE is highly dependent on QoS, because the technical performance of a system has significant impact on some dimensions of QoE (Callet, et al., 2013).

The end user satisfaction with the service directly depends on the quality and performance of network, therefore, measurement of network performance and assessment of QoS are fundamentally important. As a basic fact, an operator that provides enhanced QoS comparing to an operator that offers poor QoS has better chance to attract new customers and keep them for longer period of time (Dobrota, et al., 2012). Hence, it is essential for operators to regularly measure QoS and subsequently deal with possible challenges and address the grievances of network user(s). The measurement and analyzing process of QoS in existing heterogeneous environment is one of the challenging tasks for service providers. But the main advantage is that it enables operators to manage network in more efficient manner.

In order to measure and analyze QoS, certain parameters should be considered. Throughput, bandwidth, delay, latency, low data rate, blocked call, dropped call, packet loss rate, packet error rate, etc. are determined to be measured and analyzed as QoS parameters.

Rural and urban areas have been equally in concern about enhanced QoS of mobile network. One of the initiatives of providing of enhanced QoS to both urban and rural areas is the Digital Agenda for Europe which sets the goal to provide fast internet access to all European citizens and to take it up to 100 Mbps download rate to at least 50% of European households by 2020 (European Commission, 2010). However, the costs of deploying a fixed broadband network outside an urban area is on average higher than the cost of deploying the network in the town or village – especially when costs per household are calculated (Mason, 2008). The digital divide and deficiency of fast Internet access in rural areas still constitute an issue in the European Union (EU) member states such as the Czech Republic (Vaněk, et al., 2010).

Enhanced QoS is not only important for operators and end users, but also has impact on the worldwide ranking of communication sector of a country. Out of 167 countries, South Korea, Denmark and Iceland are on the top because of their telecom networks provide enhanced QoS. Afghanistan is placed on 156th number of the ranking list (ITU, 2015).

Mobile communication sector of Afghanistan has had tremendous growth over the last decade. Over 89% of populated area of the country is covered by telecom service; there are 6,501 installed Base Stations (BSs), 25,080,389 people have access to mobile phones, 1,910,178 are 3G broadband subscribers, 1,856,781 are internet subscribers and around 2.4 billion USD have been invested in the sector, which has resulted in long-term economic growth in the country (ATRA, 2015). So far, approximately 3,100 kilometers of optical fibers have been laid and each operator has its own high capacity microwave backbone.

Afghanistan Telecom Regulatory Authority (ATRA) collects the QoS assessment reports monthly, quarterly and annually provided by mobile operators through post-measurement tests i.e. drive test, etc. (ATRA, 2003). Until now, no research is conducted and no study is undertaken to measure and analyze QoS from either end user or technical perspective in Afghanistan. The reports collected on QoS by ATRA are not published publicly for reference purposes. Therefore, this study is also going to serve as a baseline and reference for further research on QoS in the country. The study discovers all dimensions of QoS that mobile phone users prefer. It furthermore proposes adequate technical solutions for cellular operators in order to enhance QoS and to remain competitive in the market.

The rest of the paper is organized in the following order: In *Literature review*, necessary background and related work is provided. Research methodology and materials are discussed in *Materials and methods* section. Data measurement and statistical analysis are provided in *Results and discussion* section. The adequate solutions are stated in the *Proposed technical solutions* section. The paper is compared with similar studies in *Competitive studies* section. The final part of the paper provides conclusions based on the findings.

## Literature review

There are several available studies which measure and analyze QoS of telecommunication networks i.e. regular assessment of network performance, end user survey, etc. Some of the most recent existing literatures which measure and furthermore statistically analyze QoS of mobile networks from end user perspective are described in this part. Majority of these papers provide a detailed insight into the QoS dimensions of cellular networks and recommend operators to consider satisfaction of end user as one of the most required and competitive parameters in order to remain stable in current fastest – growing and challenging technological environment.

It is stated in (Aydin and Ozer, 2005), that satisfaction of end user from an operator is associated with wide and improved network coverage, efficient customer service, enhanced QoS, and fulfilling the expectation of mobile phone user. Studies in Hong Kong (Woo and Fock, 1999), China (Wang and Lo, 2002) and South Korea (Kim, et al., 2004) have found mutual factors i.e. wide and improved network coverage, reasonable pricing policies, enhanced QoS (both voice and data), value added services and customer support.

Both (Sukumar, 2007) and (Vasundhara et al., 2016) have discovered call rates, brand image, facilitation role of front line employees, quality of network, and sale promotion packages as key dimensions of end user satisfaction.

A study has recently been conducted in Pakistan discovered that service quality, price rate, brand image, sale promotion and network coverage have significant impact on end user satisfaction (Iqbal, 2016). While, the research which has been carried out by (Khan, 2010) found that tangibles, reliability, responsiveness, convenience, assurance, empathy, and network quality have statistical significance relationship with QoS.

Recently, a number of studies for example (Sharma, 2014) in Riyadh region of Saudi Arabia, (Shefali & Riddhi, 2014) in the Ahmedabad city of India, (Khayyat & Heshmati, 2012) in the Kurdistan region of Iraq, and (Ragupathi and Prabu, 2015) has discovered the statistical significance relationship between various factors and overall satisfaction of QoS. Easy use of mobile phone, usefulness of mobile technology, enhanced mobile/data service, level of education, occupation, location, and income status are the factors which have discovered in these papers.

The QoS of mobile networks and end user satisfaction have been studied in above mentioned papers from different dimensions. But a gap of detailed study in order to measure and analyze user experience toward network, and the most unwanted situations which mobile phone users face with is missing in the literature. Therefore, the objective of this study is measurement and statistically analysis of QoS of mobile networks from end user perspective in Afghanistan and furthermore proposing of adequate technical solutions. In total, three hypotheses are tested and the relationship between categorical variables are determined. The research is furthermore compared with three similar studies and it is proved that, the paper is on one hand deeply and thoroughly covers feelings of end users from QoS and on the other hand recommends operators to deploy the most advanced and efficient schemes in their networks in order to improve QoS and to provide enhanced end to end QoE.

## Materials and methods

The survey of this research was originally conducted to study QoS and network coverage of mobile networks in Afghanistan, but primary data only related to QoS is measured and analyzed in this paper. The questionnaire (containing 15 questions) was prepared in English, Pashto and Dari Languages. All technical terms in the questionnaire were explained in such a way which were easily understandable for ordinary mobile phone users. The survey only covers end user practices, therefore, an effective evaluation method of multiple – choice questionnaire has been used.

In the beginning, authors conducted a pilot survey on a small group of mobile phone users through in – person interviews in Kabul in order to test questionnaire. Based on feedbacks from the target group of respondents, necessary change has been brought in strategy as well as confirmed the final draft of survey. Authors have subsequently employed mix – mode technique in data collection from 1,515 mobile phone users during (August – December) 2015. It specifically means, that 812 respondents were collected over the internet using Google Docs from 30 provinces and 703 respondents were interviewed in – person by volunteer surveyors within 14 specific provinces.

Total number of respondents attended the survey from all 30 provinces are shown in Figure 1. The average number of respondents per province (mean) is 50.5.

## Results and discussion

### Data measurement

All respondents have answered about their level of education, favorite mobile operator, and the purpose of mobile phone usage. Of those respondents, 1,458 were using mobile phones for telephony service, therefore, this number is

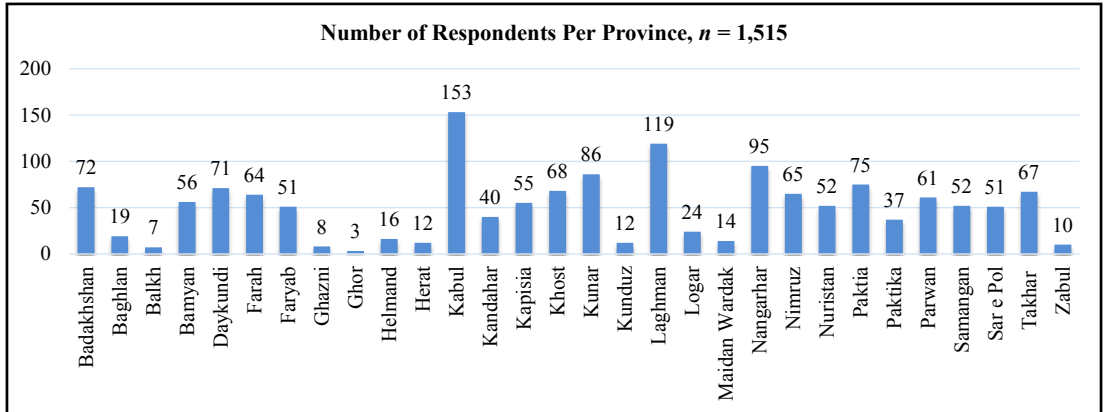

Source: Own processing

Figure 1: Number of respondents per province.

considered as random sample size for measurement of QoS and the unwanted situations of mobile telephony. 856 were using mobile internet service, thus, it is considered as random sample size for measurement and analyzing of usage of various mobile internet technologies as well as QoS and the unwanted situations related to mobile internet.

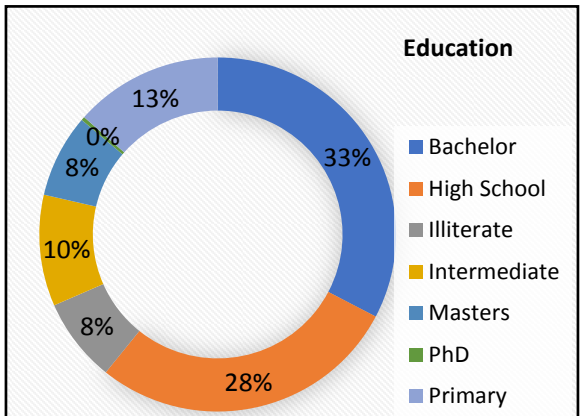

Source: Own processing

Figure 2: Level of education of mobile users.

Education has significant role in removing of barriers on the way to access/use mobile phone (Primo, 2003). Therefore, level of education of end users has been asked in the survey in order to find the importance of level of education in access/usage of mobile service in the country. Based on the result shown in Figure 2, only 8% of mobile phone users are illiterate, while, the rest are having various levels of education.

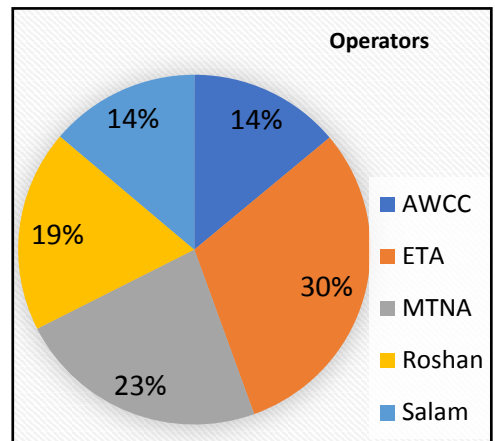

Source: Own processing

Figure 3: Mobile GSM operators.

Afghan Wireless Communication Company (AWCC), Etisalat – Afghanistan (ETA), Mobile Telecommunications Networks – Afghanistan (MTNA), Roshan and Salam are currently five GSM operators, operating inside the country. As shown in Figure 3, end users were asked about their favorite cellular operators in order to find out the leading service provider in the market.

Mobile users use a cellular phone to send/receive messages, access to the internet, send/receive email, download apps, get directions, participate in a video call, 'Check in' or share location, and so on. In the questionnaire, all of the mentioned applications of mobile phone are divided into three categories, mobile for internet purpose, mobile for telephony purpose and mobile for both internet and telephony purposes.

The result of the survey in Figure 4 shows that 48% of end users use mobile phone only for telephony service, 48% more for both internet/

telephony purposes, and only 4% of users use mobile internet in the country.

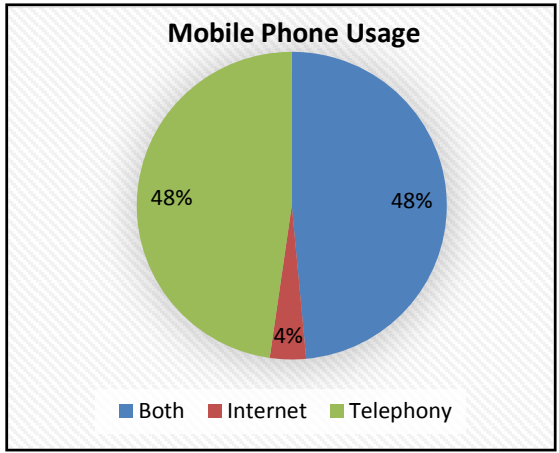

Source: Own processing

Figure 4: Mobile phone usage.

For the time being, the most advanced mobile technology which is provided by all operators in the country is UMTS. While there are still some areas which are covered by EDGE, GPRS and in some cases the end users use one of these mobile internet technologies as Wi-Fi in their homes or small offices. The result of the survey in Figure 5 shows that roughly third out of fourth of end users use 3G (UMTS) service, 14% EDGE, 7% GPRS, and 6% use mobile internet technologies for Wi-Fi purposes.

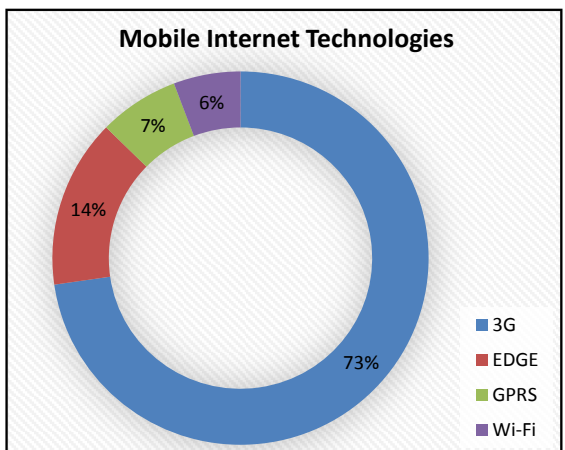

Source: Own processing

Figure 5: Mobile internet technologies.

The end users were asked about their satisfaction from QoS of mobile telephony. The result in Figure 6 shows that roughly half of the end users (45%) are satisfied, 8% very satisfied, 16% unsatisfied, 28% neutral, and 3% are very unsatisfied.

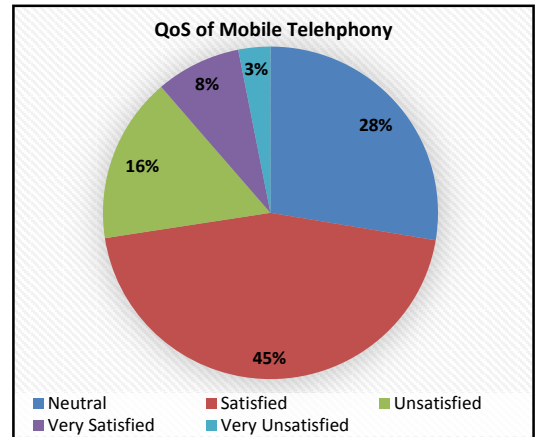

Source: Own processing

Figure 6: Satisfaction of end users from QoS of mobile telephony.

The survey explores the events which occur during/before telephony conversation and furthermore decrease the satisfaction of end users of mobile networks. The users were asked about the most unwanted situations they have been facing, i.e. blocked calls, dropped calls, echo, Low Signal Intensity (LSI) and noise. There are more other technical parameters and events, i.e. packet loss which effect on the quality of telephony service, but due to lack of technical expertise of end user they are not considered in this research.

The result of the survey shown in Figure 7 indicates that 32% of end users were complaining from LSI, 18% from blocked calls, 17% from dropped calls, 11% from echo, 14% from noise and 9% of end users were totally satisfied with QoS of mobile telephony and were not experiencing none of the above five mentioned phenomena during or before telephony conversations. The respondents were able to select only one option which they have mostly experienced.

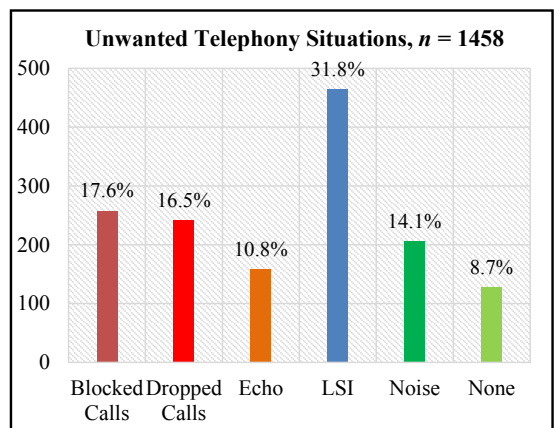

Source: Own processing

Figure 7: Unwanted mobile telephony events.

As mentioned earlier, of those 1,515 respondents, 856 were using mobile internet service, therefore, this number is going to consider as sample size for measurement and analyzing of satisfaction of end user from QoS of mobile internet as well as the events which occur while mobile internet usage. The result of the data shown in Figure – 8 declares that 22% of end users are neutral, 29% satisfied, 32% unsatisfied, 5% very satisfied, and 12% are very unsatisfied.

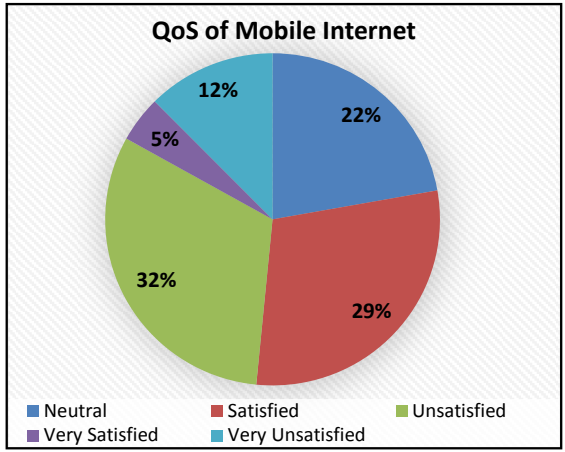

Source: Own processing

Figure 8: Satisfaction of end users from QoS of mobile internet.

The end users were further asked about the events which occur while mobile internet usage. The result in Figure 9 indicates that 33% of end users were complaining from limited coverage area of mobile internet, 27% from Low Data Rate (LDR), 13% from low mobility performance, 24% from Low Signal Strength (LSS), and 4% were not experiencing none of the above events while mobile internet usage. End users were able to select only one choice out of five for this question.

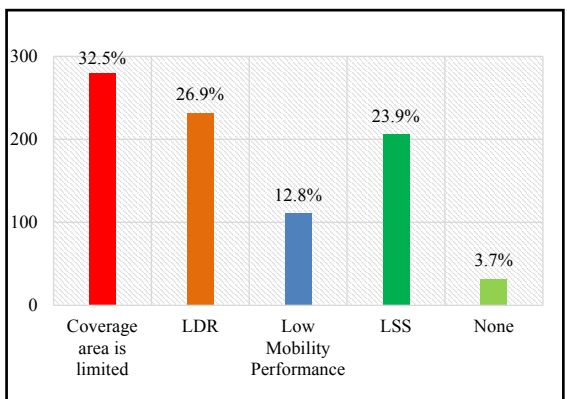

Source: Own processing

Figure 9: Unwanted mobile internet events.

### Statistical analysis

There are in total six categorical variables which create three hypotheses in collected data of this survey related to the QoS of mobile networks in Afghanistan. Associations between these variables are expected to be tested, therefore, Goodness-of-Fit (*Chi-Square*) test has been chosen to deploy on each of the hypotheses in order to find dependency between them. To test dependency, hypothesis is needed to be stated, the contingency table is expected to be created, the Degrees of Freedom (DF) is to be determined, the significance level is needed to be established, the chi – square test is to be performed, and the *Distribution table* value considering DF is compared with ch-square value respectively.

Before conducting the chi – square test, it is necessary to set up significance level. Authors consider 95% significance level ($\alpha = 0.05$) for all three hypotheses. As shown in Equation 1, it is determined by multiplying of "number of rows minus one" by "number of columns minus one".

$$DF = (r - 1) * (c - 1) \qquad (1)$$

$r$ = No. of rows, $c$ = No. of columns

In next step, the below given formula (Equation 2) is used to perform chi – square test.

$$\chi^2 = \sum_{i=1}^{k} \frac{(O_i - E_i)^2}{E_i} \qquad (2)$$

$k$ = No. of categories, = Chi-square

$i$ = No. of parameters being estimated

$O_i$ = Observed frequency, $E_i$ = Expected frequency

As mentioned earlier, after obtaining the $\chi^2$ value, it should be compared with critical value from the distribution table considering DF. The final decision is made based on this comparison. If the "chi-square value > table value", the hypothesis is rejected, otherwise, it is impossible to reject.

### First hypothesis

Is there any dependency between categorical variables of 'Education level' i.e. bachelor, high school, etc. and '*Purpose of Using of Mobile Phone*' i.e. mobile for telephony/internet/both purposes of end users of cellular networks in Afghanistan?

The first step is to state the Null hypothesis ($H_0$) and Alternative hypothesis ($H_1$).

- $H_0$ = There is no association between *'Education Level'* and *'Purpose of Using of Mobile Phone'*.
- $H_1$ = There is an association between *'Education Level'* and *'Purpose of Using of Mobile Phone'*.

The Statistical Application System (SAS) software is used in order to conduct chi – square test, create contingency table and calculate DF. Based on the results obtained from SAS, the DF and $\chi^2$ values are given below:

$\chi^2$ = 429.9240, DF = 12

The contribution table value considering 12 DF and α = 0.05 is 21.026. As calculated, contribution table value is less than comparing to chi – square test value (429.9240 > 21.026), therefore, the null hypothesis is rejected.

To conclude, there is statistically significant evidence at α = 0.05 that $H_0$ is false. Thus, it can be claimed that, there is dependency between categorical variables of *'Education level'* and *'Purpose of Using of Mobile Phone'* of end users of cellular networks in Afghanistan. It specifically means that, level of education of an end user does effect on the usage of mobile phone for different usage purposes, i.e. internet, telephony, text sending/receiving, browsing, location update, and so on.

On the other hand, (Sharma, 2014) has been investigated relationship between one of the variables of first hypothesis *'Education level'* with 'satisfaction of mobile phone user' and has found that there is significant difference between both. But, (Ragupathi and Prabu, 2015) have obtained different result and stated that, there is no significant difference between above two variables.

**Second hypothesis**

Is there any dependency between categorical variables of 'Unwanted Situations' which occurs during/before telephony conversation i.e. blocked calls, dropped calls, etc. and *Satisfaction of QoS of mobile telephony* of end user of cellular networks in Afghanistan?

The $H_0$ and $H_1$ are stated as following:

- $H_0$ = There is no dependency between *'Unwanted Situations'* and *'Satisfaction of QoS of mobile telephony'*.
- $H_1$ = There is a dependency between *'Unwanted Situations'* which occur during telephony and *'Satisfaction of QoS of mobile telephony'*.

Based on the result of SAS, the value of DF and $\chi^2$ for second hypothesis are given below:

$\chi^2$ = 307.4957, DF = 20

The contribution table value considering 20 DF and α = 0.05 is 31.410. Based on this calculation, the contribution table value is less than the chi- square test value (307.4957 > 31.410), therefore, the null hypothesis is rejected.

It is concluded, that there is statistically significant evidence at α = 0.05 that $H_0$ is rejected. It can be claimed that, there is a dependency between categorical variables of *'Unwanted Situations'* which occur during/before telephony conversation and *'Satisfaction of QoS of mobile telephony'* of end user of cellular networks in Afghanistan. It specifically means that if end users experience unwanted situations during or before telephony conversations, i.e. blocked calls, dropped calls, noise and so on are going to highly decrease their satisfaction from QoS of mobile networks.

There is no study to specifically test the relationship between above mentioned variables. However, (Khan, 2010) has been found that network quality is the most dominant dimension in affecting the customers' perception of mobile phone service quality. On the other hand, (Isabona and Ekpenyong., 2015) have identified that dropped calls and blocked calls are critical in evaluation of service quality.

**Third hypothesis**

Is there any association between categorical variables of *'Unwanted Situations'* which occur while using mobile internet i.e. low data rate, coverage area is limited, etc. and *'Satisfaction of QoS of mobile internet'* of end user of cellular networks in Afghanistan?

The $H_0$ and $H_1$ are stated below.

- $H_0$ = There is no dependency between *'Unwanted Situations'* and *'Satisfaction of QoS of mobile internet'*.
- $H_1$ = There is a dependency between *'Unwanted Situations'* and *'Satisfaction of QoS of mobile internet'*.

The value of DF and $\chi^2$ obtained by SAS for third hypothesis are given below.

DF = 16, $\chi^2$ = 139.2927

The contribution table value with α = 0.05

and considering 16 of DF is 26.296, which is less than the test value (139.2927 > 26.296). Therefore, the null hypothesis is rejected.

To sum up, based on the result of chi square test, the null hypothesis is rejected, and there is a statistically significant evidence at α = 0.05 to show dependency between categorical variables of "*Unwanted Situations*' which occur during usage of mobile internet and '*Satisfaction of QoS of mobile internet*' users in Afghanistan. It specifically means that if end users experience unwanted events during usage of mobile internet i.e. limited coverage area, low data rate, low signal intensity and so on are going to highly decrease their satisfaction from QoS of mobile networks.

Refer to the available literature and alike second hypothesis, there is no study to specifically test the relationship between above mentioned variables. Meanwhile, (Isabona and Ekpenyong., 2015) have tested that poor network coverage decrease service quality which has furthermore negative impact on end user satisfaction.

**Proposed technical solutions**

There is a couple of schemes which increase performance of a cellular network. Each scheme has its own algorithm which has advantages and disadvantages. Some of the most recent advanced schemes which are efficient to implement are discussed in this part. It is recommended for operators to deploy each of the following schemes in its appropriate location in order to enhance QoS of their networks.

Mobile networks have shifted from being predominantly voice to primarily data. In traditional networks Macro Cells (MCs) were used to cover specific geographical areas. Later on, deployment of small cells composed with macro cells has emerged, which improves cellular coverage, capacity and applications for homes and enterprises as well as metropolitan and rural public spaces. Increased number of small cells in a given zone enhances performance and theoretically serves higher number of users, but many challenges i.e. interference, mobility management, radio resource management and so on are appeared. As shown in Figure 10, $MC_1$ covers a specific zone. But, later on small cells ($SC_{1, 2, 3, 4, 5, 6, 7 \text{ and } 8}$) are added to the topology in order improve network coverage and enhance QoS.

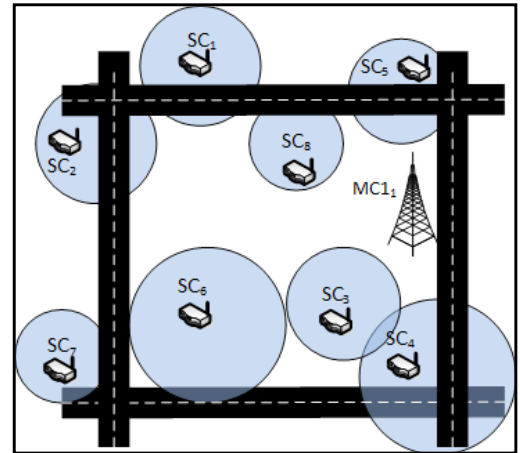

Source: Own processing

Figure 10: Deployment of small cells.

Small cells have short coverage area, therefore, it is recommended to deploy this scheme mostly in urban areas, homes, and enterprise where the density of traffic is high. Moreover, high care should be taken while selecting the position and establishing transmitting power of small base stations which have significant impact on network performance (Kelif, et al., 2013).

High order sectorization is the second scheme, which is recommended for operators in order to increases capacity and enhances QoS, but cell interference and number of softer – handovers are main challenges which rise with deployment of this scheme. Traditional cellular operators use macro cells with wide beam antennas for wider coverage, but high demand capacity of end users in existing and emerging networks cannot be achieved by deployment of only macro cells (Sheikh and Lempiainen, 2013). Therefore, it is required to achieve maximum practical capacity from macro cells by employing higher order sectorization and by utilizing all possible antenna solutions including smart antennas. As shown in Figure 11 (*a*), coverage area of a cell is divided into three sectors (3*120°). It is noted that cloverleaf layout is offered the lowest interference level which provides best cell and system capacity for macro cells. However, this scheme is not efficient for high density area and current demand of end user, therefore, maximum capacity utilization of macro cells is guaranteed by increasing of number of sectorization as shown in *b* (4*90°), *c* (6*60°), and *d* (8*45°) of Figure 11.

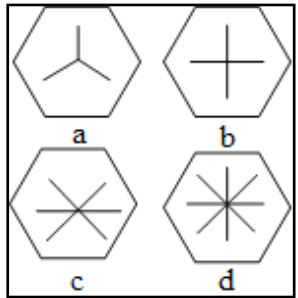

Source: Own processing

Figure 11: High order sectorization.

The placement of both macro and small base stations is a challenging task due to its interference constraints with other cells and proper coverage of a specific area where all users access to enhanced QoS. A method of optimal placement of femto cell in order to increase the QoS in dense environment where macro-cell holds many number of femto cells is discussed by (Kilaru and Gali, 2015). The proposed scheme made an assumption that the interference effect is considerably strong compared to noise. The result of the simulation has shown that the optimal placement has better throughput compare to blind placement in the topology. As shown in Figure - 12, a large area is remained uncovered due to huge amount of blackpost which is caused by wrong selection of tower location. Therefore, optimum location of base station has significant impact on enhancing of QoS.

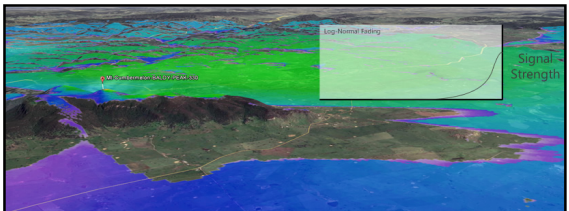

Source: Telecoantenna, 2015

Figure 12: Optimum location of base station.

Regular post – performance measurement tests i.e. drive test for optimization purpose is a traditional scheme, which is used by all cellular operators. This scheme helps to find existing gaps in network coverage and measure QoS. But, high number of these tests will lead to spend more time, money and personnel in order to collect relevant data and furthermore analyze the problems (Isabona and Obahiagbon, 2014). Therefore, it is necessary to have optimum number of test per time period.

Increasing number of carriers (TRXs) and base station parameters i.e. transmitting power can also be useful schemes to increase QoS considering geographical location and population of the area (Haider, et al., 2009). These methods are helpful to increase the capacity of a base station, which leads to enhance QoE. As far as base station can support a limited number of TRXs, therefore, this scheme will be more efficient if it is combined with one of the mentioned proposed solutions. However, it is crucial to remember that increasing of values of parameters from defined standards rise additional challenges e.g. environmental effect and so on.

As discussed, all proposed technical schemes have its own characteristics – pros and cons. But appropriate time and place of deployment of these schemes require detailed study of geographical location, population of the area, and government regulatory policies.

**Competitive studies**

Three research papers are compared with this study. The first is exploratory study, performed by (Khayyat and Heshmati, 2012) which discusses mobile phone user satisfaction in the Kurdistan region of Iraq. The second is empirical research conducted by (Khan, 2010) and the third paper analyzes QoS of mobile networks in Pakistan (Iqbal, 2016).

All these papers measured and analyzed QoS and satisfaction of mobile phone users from end user perspective, but have three main drawbacks. The first one is inclusion of some unnecessary parameters and testing of unrelated hypothesis. For example, (Khayyat and Heshmati, 2012) has tested the relationship between age, gender and occupation with satisfaction of end user from QoS. These parameters and hypotheses do not have any impact on QoS of mobile networks, thus, there is no need to test. The second weak point is the exclusion of recommended solutions. Proposing of adequate solutions of author(s) is a major part of a research paper and should be highly considered. The third one is less amount of sample size. Considering geographical location as well as the population of each of area where above three studies were conducted, it would be more efficient if authors collected high number samples to address the issue more accurately.

Therefore, all existing weaknesses in current state of the art were considered and furthermore thoroughly addressed in this paper. Therefore, the strengths of this research are not only in specific measurement and detailed analysis of QoS but also inclusion of recommended technical solutions for operators and considering higher number of sample size than all competitive papers so far.

# Conclusion

In this paper, authors measured and statistically analyzed QoS of mobile networks from end user perspective in Afghanistan. A survey was conducted from 1,515 mobile phone users of five cellular operators. In total three hypotheses are tested and the relationship between specific categorical variables have been determined. In order to address existing challenges in the area of QoS of mobile networks in Afghanistan, authors recommended deployment of small cells, increasing number of regular performance tests, optimal placement of base stations, increasing number of carriers, and high order sectorization as adequate technical solutions. Comparison of this research with three similar studies has shown that unlike other papers the research on one hand deeply and thoroughly covered feelings of end users from QoS and on the other hand recommended operators to deploy the most advanced and efficient schemes in their networks. In future work, authors intend to focus on measurement and analysis of QoS of mobile networks from network perspective in Afghanistan.

# Acknowledgements

The results and knowledge included herein have been obtained owing to support from the Internal grant agency of the Faculty of Economics and Management, Czech University of Life Sciences in Prague, grant no. 20161012 "Potential of the use of Internet of Things impacting regional development and agrarian sector" (In Czech: "Potenciál využití Internet of Things s důrazem na rozvoj regionů a agrárního sektoru").

*Corresponding author:*
*Mohammad Asif Habibi*
*Department of Information Technologies, Faculty of Economics and Management*
*Czech University of Life Sciences Prague, Kamýcká 129, Prague 6 - Suchdol, 165 21, Czech Republic*
*Phone: +420 776 716 837, E-mail: masif.habibi1988@gmail.com*